\documentclass[11pt]{article}
\usepackage[margin=1.0in]{geometry}
\usepackage{graphicx}
\usepackage{epsfig}
\usepackage{amssymb,amsfonts}
\usepackage{mathrsfs}
\usepackage{epstopdf}
\usepackage{hyperref,color}
\usepackage{natbib}
\usepackage{setspace}
\usepackage{xypic}\usepackage[all,cmtip]{xy}

\pdfoutput=1

\begin{document}

\title{No categorial support for radical ontic structural realism} 
\author{Vincent Lam and Christian W\"uthrich\thanks{This work is fully collaborative. We wish to thank the audience at the meeting of the European Philosophy of Science Association in Athens, the philosophy of physics group at UCSD, and the referees of this journal, Jonathan Bain, Craig Callender, Erik Curiel, Michael Esfeld, Oliver Pooley, and particularly Elias Zafiris for correspondence and comments. C.W.\ acknowledges support from the American Council of Learned Societies through a Collaborative Research Fellowship, the University of California through a UC President's Fellowship in the Humanities, and the University of California, San Diego, through an Arts and Humanities Initiative Award. V.L. is grateful to the Australian Research Council (ARC, Discovery Early Career Researcher Award, project DE120102308)  and to the Swiss National Science Foundation (SNSF, Ambizione grant PZ00P1\_142536/1) for financial support.}}
\date{}
\maketitle

\begin{abstract}\noindent
Radical ontic structural realism (ROSR) asserts an ontological commitment to `free-standing' physical structures understood solely in terms of fundamental relations, without any recourse to relata which stand in these relations. \citet{Bain2011} has recently defended ROSR against the common charge of incoherence by arguing that a reformulation of fundamental physical theories in category-theoretic terms (rather than the usual set-theoretic ones) offers a coherent and precise articulation of the commitments accepted by ROSR. In this essay, we argue that category theory does not offer a more hospitable environment to ROSR than set theory. We also show that the application of category-theoretic tools to topological quantum field theory and to algebraic generalisations of general relativity do not  warrant the claim that these theories describe `object-free' structures. We conclude that category theory offers little if any comfort to ROSR.
\end{abstract}

\begin{center}
\end{center}

\section{Introduction: ridding structures of objects}
\label{sec:intro}

Eliminative or radical ontic structural realism (ROSR) offers a radical cure---appropriate given its name---to what it perceives to be the ailing of traditional, object-based realist interpretations of fundamental theories in physics: rid their ontologies entirely of objects. The world does not, according to this view, consist of fundamental objects, which may or may not be individuals with a well-defined intrinsic identity, but instead of physical structures that are purely relational in the sense of networks of `free-standing' physical relations without relata.\footnote{This radical or eliminative version of OSR has been defended more or less explicitly in \cite{French1998}, \cite{FrenchandLadyman2003} and \cite{Saunders2003b}, and as recently as in \cite{French2010}, among others.} 

Advocates of ROSR have taken at least three distinct issues in fundamental physics to support their case.\footnote{There are actually two main groups of motivations for ROSR that can be found in the literature (see for instance \citealp[\S\S II-III]{French2006}). The first concerns the standard challenge to scientific realism from theory change (in the recent debate this challenge indeed constitutes a motivation for structural realism in general, not only its ontic version). The second concerns the ontological implications of fundamental physics, such as the issues we mention here.} The quantum statistical features of an ensemble of elementary quantum particles of the same kind as well as the features of entangled elementary quantum (field) systems as illustrated in the violation of Bell-type inequalities challenge the standard understanding of the identity and individuality of fundamental physical objects: considered on their own, an elementary quantum particle part of the above mentioned ensemble or an entangled elementary quantum system (that is, an elementary quantum system standing in a quantum entanglement relation) cannot be said to satisfy genuine and empirically meaningful identity conditions. Thirdly, it has been argued that one of the consequences of the diffeomorphism invariance and background independence found in general relativity (GR) is that spacetime points should not be considered as traditional objects possessing some haecceity, i.e.\ some identity on their own. 

The trouble with ROSR is that its main assertion appears squarely incoherent: insofar as relations can be exemplified, they can only be exemplified by some relata. Given this conceptual dependence of relations upon relata, any contention that relations can exist floating freely from some objects that {\em stand in} those relations seems incoherent. If we accept an ontological commitment e.g.\ to universals, we may well be able to affirm that relations exist independently of relata---as abstracta in a Platonic heaven. The trouble is that ROSR is supposed to be a form of scientific realism, and as such committed to asserting that at least certain elements of the relevant theories of fundamental physics faithfully capture elements of physical reality. Thus, a defender of ROSR must claim that, fundamentally, relations-sans-relata are exemplified in the physical world, and {\em that} contravenes both the intuitive and the usual technical conceptualization of relations.

The usual extensional understanding of $n$-ary relations just equates them with subsets of the $n$-fold Cartesian product of the set of elementary objects assumed to figure in the relevant ontology over which the relation is defined. This extensional, ultimately set-theoretic, conceptualization of relations pervades philosophy and operates in the background of fundamental physical theories as they are usually formulated, as well as their philosophical appraisal in the structuralist literature. We will have to come back to this point, but the charge then is that the fundamental physical structures that are represented in the fundamental physical theories are just not of the `object-free' type suggested by ROSR.

While ROSR should not be held to the conceptual standards dictated by the metaphysical prejudices it denies, giving up the set-theoretical framework and the ineliminable reference to objects and relata attending its characterizations of relations and structure requires an alternative conceptualization of these notions so central to the position.\footnote{\citet[52n]{FrenchandLadyman2003} are well aware of this difficulty: ``If we're going to take our structuralism seriously, we should therefore be appropriately reflective and come up with thorough-going structural alternatives to group theory and set-theory.'' More recently, however, \citet[23]{french2012} seems to have moved away from this position and now thinks that ROSR is compatible with a set-theoretic understanding of structure, as long as the relations take ontological precedence over the relata. That, we maintain, cannot succeed, as the articulation of a structure so understood ineliminably contains objects.} This alternative conceptualization remains necessary even in the light of `metaphysics first' complaints, which insist that ROSR's problem must be confronted, first and foremost, at the metaphysical level, and that the question of how to represent structure in our language and in our theories only arises in the wake of a coherent metaphysical solution. But the radical may do as much metaphysics as she likes, articulate her theory and her realist commitments she must, and in order to do that, a coherent conceptualization of what it is to have free-floating relations exemplified in the physical world is necessary. 

ROSR thus confronts a dilemma: either soften to a more moderate structural realist position or else develop the requisite alternative conceptualizations of relations and of structures and apply them to fundamental physical theories. A number of structural realists have grabbed the first leg and proposed less radical and non-eliminative versions of ontic structural realism (OSR).\footnote{See for instance \citet[Ch.~3]{LadymanRossSpurrettandCollier2007}, \citet{Ladyman2007}, and \citet{EsfeldandLam2008}.} These moderate cousins of ROSR aim to take seriously the difficulties of the traditional metaphysics of objects for understanding fundamental physics while avoiding these major objections against ROSR by keeping some thin notion of object. The picture typically offered is that of a balance between relations and their relata, coupled to an insistence that these relata do not possess their identity intrinsically, but only by virtue of occupying a relational position in a structural complex. Because it strikes this ontological balance, we term this moderate version of OSR `balanced ontic structural realism' (BOSR).

But holding their ground may reward the ROSRer with certain advantages over its moderate competitors. First, were the complete elimination of relata to succeed, then structural realism would not confront any of the known headaches concerning the {\em identity} of these objects or, relatedly, the status of the Principle of the Identity of Indiscernibles; in particular, the embarrassment rearing in highly symmetrical situations such as the Friedmann-Lema\^{\i}tre-Robertson-Walker spacetimes in GR \citep{Wuthrich2009} or an assembly of elementary quantum particles---especially bosons---would be circumvented. To be sure, this embarrassment can arguably be avoided by other moves;\footnote{For instance, one could invoke a weaker version of the Principle of the Identity of Indiscernibles, see \citet{Muller2011} in the GR case and \citet{MullerandSaunders2008} and \citet{MullerandSeevinck2009} in the quantum statistical case, although whether this strategy succeeds is debatable.} but eliminating objects altogether simply obliterates any concerns whether two objects are one and the same. Secondly, and speculatively, alternative formulations of our fundamental physical theories may shed light on a path toward a quantum theory of gravity. 

For these presumed advantages to come to bear, however, the possibility of a precise formulation of the notion of `free-standing' (or `object-free') structure, in the sense of a network of relations without relata (without objects) must thus be achieved. Recently, Jonathan \citet{Bain2011} has argued that category theory provides the appropriate mathematical framework for ROSR, allowing for an `object-free' notion of relation, and hence of structure. This argument can only succeed, however, if the category-theoretical formulation of (some of the) fundamental physical theories has some physical salience that the set-theoretical formulation lacks, or proves to be preferable {\em qua formulation of a physical theory} in some other way. This argument from category theory in favour of ROSR is the focus of this paper. 

We assert that the question of the possibility of some precise mathematical representation of the notion of an `object-free' structure is important and interesting on its own right, independently of one's stance on ROSR. Indeed, this question touches foundational issues such as the role and status of category theory in formulating and developing fundamental physical theories (such as in quantum gravity) and the relationship between category theory and set theory. Furthermore, it contributes to the ongoing discussion among structuralists as to the best characterization of structure.\footnote{Cf.\ e.g.\ \citet{Landry2007,Muller2010}.}

F.~A.~\citet{Muller2010} has argued that neither set theory nor category theory provide the tools necessary to clarify the ``Central Claim'' of structural realism that the world, or parts of the world, {\em have} or {\em are} some structure. The main reason for this arises from the failure of reference in the contexts of both set theory and category theory, at least if some minimal realist constraints are imposed on how reference can function. Consequently, Muller argues that an appropriately {\em realist} stucturalist is better served by fixing the concept of structure by axiomatization rather than by (set-theoretical or category-theoretical) definition. Although we would not want to be bound by his strong claims that no realist story of the reference of set- or category-theoretically defined structure cannot be given, we agree that `going native' by axiomatizing the concept of structure is desirable, or may ultimately even be necessary. Our general project, however, is to argue that category theory offers no better environment for ROSR than does set theory.\footnote{We would like to emphasise that our main focus is \citet{Bain2011}'s argument in favour of category theory as an appropriate framework for the `object-free' physical ontology suggested by ROSR; in particular, we will not directly address the issue of the precise relationship between mathematical formalism and physical ontology---we take it for granted as Bain does.}

In Section \ref{sec:peril}, we explicate the various aspects of the threat to ROSR and tabulate the tasks the ROSRer must complete in order to avert the threat. Section \ref{sec:bain} presents Bain's recent attempt to save ROSR based on a reformulation of fundamental physical theories based on category theory, introducing the relevant notions along the way. The next section, \S\ref{sec:throwing}, addresses two worries concerning the conceptualization of relations and their relata in a category-theoretic setting. This leads to more general considerations---the topic of Section \ref{sec:structures}---concerning the contrast between categorial and set-theoretical notions of structures. Finally, Sections \ref{sec:TQFT} and \ref{sec:einalg} discuss, in relevant detail, Bain's proposal of how quantum field theory and general relativity, respectively, can be reformulated in terms of category theory. Conclusions follow in Section \ref{sec:conclusions}.

\section{The set-theoretic peril for ROSR}
\label{sec:peril}

Objections against the radical `object-free' metaphysics of ROSR can be catalogued into three types: 
\begin{itemize}
\item[(1)] \emph{Metaphysical objection.} Concrete physical relations instantiated in the physical world require something concrete and physical that stands in the relations. Physical relations are conceptually and ontologically dependent on there being physical relata to be related (even though according to the structuralist the identity of these relata is similarly dependent upon the relations). As a consequence, ROSR cannot be what it purports to be: a metaphysical thesis about the concrete physical world.
\item[(2)] \emph{Logical or set-theoretic objection.} The irreducible quantification over objects in standard first-order logic and more generally the irreducible reference to the elements of the sets used in the mathematical formulations of physical theories imply that it is vain to argue in favour of ROSR on the basis of physical theories that are all mathematically formulated in ultimately set-theoretic terms.
\item[(3)] \emph{Physical objection.} It is not clear that current fundamental physics really requires that the notion of fundamental physical object should be eliminated altogether at the fundamental level, and not merely revised. Despite the difficulties mentioned in the introduction, quantum theory can still be understood as making reference to some fundamental objects, the nature of which is much debated, precisely because the standard notion of object has to be revised---not eliminated. Similarly, physical models of GR still make fundamental reference to spacetime points. Thus, regardless of questions of formalisms of the kind covered by the second objection above, the correct interpretation of particular physical theories does not abandon relata altogether.
\end{itemize}
(1) is perhaps the first obvious objection to ROSR that comes to mind. While this metaphysical objection does get some traction intuitively, it is open to some straightforward counters by ROSR proponents. For one thing, it relies on some distinction between mathematical and physical structures that can be rejected by the radical (see for instance \citealp[\S 3.6]{LadymanRossSpurrettandCollier2007}). More generally, it can be argued that invoking some sort of mutual conceptual and ontological dependence of physical relations and physical relata against ROSR begs the question precisely because ROSR denies such dependence \citep[871]{Chakravartty2003}. Given her insistence that our metaphysics be scientifically informed, the radical can argue that our naturalized metaphysics is subject to sometimes quite deep and counterintuitive revision, just as is our science. Again, of course, simply denying the terms on which (1) rests will not suffice---an alternative conception must be worked out.

One way to attempt such a conception would be to articulate a `bundle-theoretic' approach to ROSR. Fundamentally, on this approach, and conducive to ROSR, there are only relations and no relata. Physical objects, on this view, are ontologically dependent upon the relations they exemplify; in fact, they are constituted by their bundle of compresent relations. On this non-eliminativist bundle version of ROSR, the relata would still exist, albeit not independently. Naturally, one of the primary tasks of bundle-theoretic ROSR is to give an account of the compresence of the relations exemplified by, and `bundled' in, the dependent object which does not tacitly rely on there being relata. In particular, compresence must be explicated in a way that would not lend itself to straightforward reinterpretation in terms of metaphysically thin objects of the kind accepted by BOSR. 

The set-theoretic objection (2), unlike the metaphysical objection (1), cannot be dismissed on the ground that it is metaphysically prejudiced. (2) is based on formal, albeit simple, considerations from set theory. Indeed, the standard way of modeling the extensional understanding of the central ROSR notions of relation and structure in set theory makes ineliminable reference to the elements of the sets involved in the definitions. A binary relation on a set $A$ is defined as a set of ordered pairs $\langle a_1, a_2\rangle$, where $a_1, a_2 \in A$ and where the ordered pair $\langle a_1, a_2\rangle$ can be defined to be the set $\{a_1, \{a_1, a_2\}\}$. More generally, an $n$-ary relation on sets $A_1,..., A_n$ is a subset of the Cartesian product $A_1 \times\cdots\times A_n$ of these sets, where the Cartesian product is defined as the set of all ordered $n$-tuples $\langle a_1,..., a_n\rangle$ such that for all $i=1,...,n$, $a_i\in A_i$. Thus, this set-theoretical definition of relations makes irreducible reference to the elements of the set on which they are defined. A set $A$ of elements together with a set $R$ of relations defined on it constitute the standard set-theoretical characterization of the notion of a (relational) structure, which can be merely defined as the ordered pair $\langle A, R\rangle$.\footnote{A more rigorous definition of `structure' can be found in \citet[\S 1]{Wuthrich2011}.} This set-theoretical definition of structure clearly also makes ineliminable reference to the elements of $A$ (sometimes called the `domain' of the structure).

More generally, the very fact that sets are entirely and solely characterized by their elements (as encoded in the axiom of extensionality within the Zermelo-Fraenkel axiomatization) can be invoked in order to argue that reference to the elements of the sets is unavoidable. To the extent that these set elements can be understood as representing possible physical objects (and relata, as in the case of the definition of a relation), set theory then seems to constitute a rather hostile environment for ROSR. One may resist the inference from the fact that a set-theoretic formulation requires a domain of quantification to the conclusion that the elements in this domain must be the relata ROSR seeks to eliminate. The fact that one quantifies over a domain of `objects' does, by itself, entail nothing about the nature of these `objects': these `objects' may well be properties or relations! 

We see no reason to think that a `dual' (but still set-theoretic) characterization of structure in which we start from a domain of relations on which we then define objects, i.e.\ entities which represent the relata outmoded by ROSR, to obtain the full structure cannot succeed. What we mean by a `dual' characterization of structure can be made salient by considering a graph-theoretical representation of the standard extensional understanding of structure in which we have `vertices'---the relata---connected by `edges'---the relations. A dual of a graph-theoretically characterized structure would then simply be the graph-theoretical structure obtained by mapping the vertices to edges and vice versa; instead of two vertices `connected' by an edge, we would then have a vertex `binding' two edges. Such `dual' characterizations, however, would still require a two-sorted conceptualization of structure including, e.g., domains and `objects', or edges and vertices. Insofar as the set-theoretic conception requires both sorts to characterize structures, it needs something more than just an unstructured set of relations. None of this suggests a need to return to physical objects as traditionally conceived, of course; but, the objection states, even if it can be shown how to formulate physical theories such that there are only relations in the domain, any ultimately set-theoretic conception of structure must be two-sorted.\footnote{We thank a referee for pressing us on this point.}

In a way, this is the representational analogue of the difficulty that we can find e.g.\ in bundle-theoretic ROSR, where it proved insufficient to just state that fundamentally, all there is are relations. The bundle theorist seems to require the metaphysical counterpart of a two-sorted conception in the representation of structure: instead of edges and vertices, she needs relations as well as bundles. 

Before we proceed, however, we would like to emphasize that for our present purposes, not much hangs on whether an extensional, ultimately set-theoretical conception of relations and of structure can be embedded into a more hospitable framework for ROSR. Perhaps it can. The relevant point, however, is that the main target of this essay, the argument by Bain that we are about to introduce, accepts this standard objection against ROSR and aims to address it head-on. 

Quite regardless of concerns regarding general metaphysical matters or the formal representation of physical structure, objection (3) expresses the worry that the correct way of interpreting our best physical theories does not, as ROSR claims, strongly suggest, let alone necessitate, the complete elimination of physical objects. 
In other words, there is no motivation for ROSR coming directly from the physics of quantum systems or of spacetime.\footnote{The issue of the (lack of) motivation for ROSR is discussed in \citet{Chakravartty2003} and \citet{Ainsworth2010}.} It is important to realize that this complaint against ROSR is consistent with the recognition that modern physics lays bare the fact that traditional ontological categories such as that of a physical object are inadequate. What is needed, but not given, according to this demurral, is an argument to the effect that ROSR is superior to BOSR as a template to interpret fundamental physical theories. 

\cite{Bain2011} argues that category theory offers an alternative and more hospitable setting for ROSR, allowing ROSR to resist, at least to some extent, the above objections. Before turning to the details of his argument in the next section, we insist that the following four tasks have to be completed in order for ROSR to clinch victory, as least as envisaged by Bain. The first two concern the mathematical formalism, and only the third and fourth tasks apply this formalism to the physical world.  (i) First, it has to be shown that category theory---contrary to set theory---does indeed allow for notions of `free-standing' relations and structure that do not rely on some notions of object and relata. (ii) For the previous claim to gain strength, it has to be shown that these category-theoretic notions of `free-standing' relations cannot be isomorphically reformulated in set-theoretical language, otherwise the worry would arise that such categorial move amounts to nothing else but mere relabelling. (iii) Since ROSR is a thesis about the physical world as described by fundamental physics, it has to be shown that fundamental physics can be reformulated in such category-theoretic terms, and within those categories without set-theoretic equivalent. (iv) For such category-theoretic reformulation to be preferred over the set-theoretic one, it has to be shown that the category-theoretic reformulation of fundamental physics is somehow superior to the usual formulations involving set-theoretic notions, and in particular a set-theoretic understanding of `structure' and `relation' (in other terms, it has to have some physical significance that the set-theoretic formulation lacks).

To be sure, an account satisfying (i) through (iii) would already be of considerable interest, even if the superiority of the category-theoretical approach would not thereby be established. But it would amount to the significant result of a possibility proof of a thoroughgoing category-theoretical reformulation of fundamental physical theories. Conversely, one might think that for ROSR to win, all it would take would be to establish an advantage at the interpretational level. According to this thought, it would be sufficient to show that there exist equivalent formulations of the physical theories at stake in purely categorial terms, with a native categorial understanding of `free-standing' relations, and that these categorial formulations are to be preferred over the standard ones. In other words, the task list would be as above, except that the last clause of task (iii) is deleted. Again, it would be a significant achievement to discharge this slightly lesser task list. But then the worry---all too familiar to structuralists of all stripes---would be that given that both the standard and the categorial formulations are {\em equivalent}, i.e., they may well express differently what is precisely the same theory, there could be nothing that would underwrite a principled metaphysical argument to the uncontested preference of one over the other. If, on the other hand, the two are not equivalent, and it can be shown that one is to be preferred over the other, this worry gets no traction. 

Futhermore, steps (iii) and (iv) would arguably go toward addressing objection (3) above. Admittedly, objections of this type crucially depend on the feasibility of interpretations of fundamental physical theories that do not deny the existence of objects. Of course these alternatives exist, both within and without the structuralist camp. To what extent they solve the interpretational worries we mentioned at the outset is subject to debate. But if the ROSRer wishes to deny that more moderate versions of OSR such as BOSR resolve these difficulties, then she must accept the burden to show as much. Burden-of-proof arguments are notorious, of course, and we accept that the ROSRer may similarly insist that the BOSRer must shoulder the prior burden of showing that their approach works. We wish we had the luxury of being able to choose between two fully workable metaphysical approaches to fundamental physics; alas, none of the tasks (i) through (iv) is trivial. 

Bain hopes to achieve outright victory by trying to establish both the superiority of theories formulated in category-theoretical terms as well as that these theories must be interpreted in the sense of ROSR; i.e., they {\em cannot} be understood in the sense of BOSR, let alone in non-structuralist terms. It is important to note that ROSR can win with quite a bit less: it is by no means necessary to show that fundamental physical theories cannot be cast in a way that admits of an interpretation conducive to BOSR or even to the non-structuralist. What advocates of ROSR need to show is thus not the {\em necessity} of their interpretation, but rather their preferability over alternatives. Thus, Bain's argument, if successful, establishes rather more than it needs to. With this point in mind, let us now turn to the details of this argument.

\section{Bain's categorial strategy to save ROSR}
\label{sec:bain}

The first step of the strategy deployed in \cite{Bain2011} is to use standard category-theoretic tools to redefine the notions of relation and structure in a way that avoids ineliminable reference to relata and objects. Category theory with its emphasis on `mappings' (called `morphisms' or `arrows') between the elements of a collection seems to be the promising framework. Indeed, one of the general themes of category theory can be described as follows: ``Instead of defining properties of a collection by references to its members, [...] one can proceed by reference to its \emph{external} relationships with other collections. The links between collections are provided by functions, and the axioms for a category derive from the properties of functions under composition.'' \cite[1, cf.\ also 37]{Goldblatt1984} We now have to introduce a few definitions in order to implement and discuss this strategy. 

A category $\mathbf{C}$ consists of a pair of classes, the class $(A, B, C,...)$ of $\mathbf{C}$-objects and the class $(f, g, h,...)$ of $\mathbf{C}$-morphisms, together with
\begin{itemize}
\item an assignment to each $\mathbf{C}$-morphism $f$ of a pair of $\mathbf{C}$-objects dom$f$ and cod$f$ (if $A =$ dom$f$, $B=$ cod$f$, one writes $f: A \rightarrow B$), 
\item an assignment to each pair of $\mathbf{C}$-morphisms $f, g$ with dom$g =$ cod$f$ of a $\mathbf{C}$-morphism (`composition') $g \circ f:$ dom$f \rightarrow$ cod$g$ such that the associative law is satisfied, i.e. $h \circ (g \circ f) = (h \circ g) \circ f$, where the $\mathbf{C}$-morphism $h$ is such that dom$h =$ cod$g$, and finally 
\item an assignment to each $\mathbf{C}$-object $B$ of an identity morphism $1_B : B \rightarrow B$ such that for any $\mathbf{C}$-morphisms $f: A \rightarrow B$ and $g: B \rightarrow C$, we have $1_B \circ f = f$ and $g \circ 1_B = g$.
\end{itemize}

Identity morphisms highlight the importance of morphisms in the definition of a category. Indeed, it is easy to see that to each object in a category corresponds exactly one identity morphism and vice versa, so that it is possible to reformulate in an equivalent way the above definition of a category only in terms of morphisms, i.e.\ without any explicit reference to the objects. In this sense, the concept of a category need not be two-sorted. Bain's argument does not, however, rely on this `object-free' redefinition of a category, for two reasons. First, identifying the objects of a category with their corresponding identity morphisms does not eliminate these objects, but can instead be considered a mere relabelling.\footnote{\label{fn:morphonly}In parenthesis, we would like to add that the complete data for a morphisms-only category consists of the morphisms, certain ordered pairs of morphisms, as well as an operation called `composite' \citep{maclane98}. It is thus clear that the morphisms come with some additional structure, which must be articulated in a non-question-begging way. Whether or not that can be achieved in purely categorial terms and without recourse to the objects of a category is subject to debate.} Second, the objects of a category (in the technical category-theoretic sense defined above, i.e.\ $\mathbf{C}$-objects) are not the focus of Bain's argument since they are actually not the possible representators of the fundamental physical objects that ROSR wants to eliminate. In a rough analogy, the category as a whole articulates the theory, and its objects represent models of the theory, not physical objects in their own right. Thus, the idea is rather to consider structures themselves as objects in a category, i.e.\ as $\mathbf{C}$-objects.\footnote{That a structure as a whole can be considered as an object (a $\mathbf{C}$-object indeed) does not pose any problem for ROSR, whose claim is that there are no fundamental, ultimate objects and relata. In order to avoid any confusion, we wish to highlight the distinction between the technical category-theoretic notion of an object ($\mathbf{C}$-object) and the notion of a fundamental physical object, understood for instance as what enters into physical relations; obviously, the ROSR proponent only aims to get rid of the latter.} Now, from this perspective we see that what ROSR actually wants to eliminate from its ontology are the {\em elements} of the objects of such a category, that is, the elements of the $\mathbf{C}$-objects for some category $\mathbf{C}$, where the notion of an element of a $\mathbf{C}$-object can be precisely defined within the category-theoretical framework (see below). These elements are the natural candidates for being interpreted as representing physical objects. Bain crucially points out that such a categorial definition allows one to eliminate explicit reference to the elements of the objects of such category (that is, to the elements of the $\mathbf{C}$-objects), and it is for this reason that category theory is supposed to help ROSR.

Instead of making (futile) use of an `object-free' conception of categories, therefore, Bain's strategy is to look at the category $\mathbf{Set}$, whose objects are sets and whose morphisms are functions between sets, and to reformulate the set-theoretic notion of an element of a set (that is, the elements of a $\mathbf{Set}$-object) in categorial terms. From $\mathbf{Set}$, other categories can be obtained by imposing some structure on the objects, i.e.\ on the sets, and by demanding that the functions between the sets leave these structures invariant. The claim then is that the set-theoretic notion of an element of a set (and more generally the notion of an element of an object of a category) becomes harmless to ROSR at the categorial level. Either the objects of a category at stake relevantly resemble the $\mathbf{Set}$-objects, i.e.\ sets, or they don't. If they do, a defence of ROSR requires that it be shown that the elements of these sets do not threaten to reverse ROSR's eviction of all objects from its fundamental ontology. If they don't, all that is left to prove is that the structure of the category's objects is benevolent to ROSR and its abhorrence of fundamental objects. Let's look more closely at the two cases in turn.

First, in categories `sufficiently similar' to $\mathbf{Set}$, where the notion of an element of an object can be defined, it is `externalized' (or `structuralized') in terms of morphisms. This can be achieved using the notion of a `terminal object' in a category $\mathbf{C}$, which is defined as the $\mathbf{C}$-object $1$ such that for every other $\mathbf{C}$-object $A$ there is one and only one $\mathbf{C}$-morphism $A \rightarrow 1$.\footnote{\label{fn:isom}The terminal object $1$ of a category should not be confused with the identity morphism $1_A$ of an object $A$ of the category. We will sometimes speak of `the' terminal object of a category $\mathbf{C}$ since for any terminal object $1$, the only morphism $1\rightarrow 1$ is the identity, and any two terminal objects $1$ and $1'$ of $\mathbf{C}$ are isomorphic in $\mathbf{C}$, as can easily be seen. Two objects $A$ and $B$ of a category $\mathbf{C}$ are {\em isomorphic} just in case there exists a morphism $f: A\rightarrow B$ in $\mathbf{C}$ such that there is a morphism $g: B\rightarrow A$ in $\mathbf{C}$ which satisfies $f\circ g = 1_B$ and $g\circ f = 1_A$. Since  $1'$ is a terminal object, there is a (unique) morphism $f: 1\rightarrow 1'$; and since $1$ is a terminal object, there is a (unique) morphism $g:1'\rightarrow 1$. But if $1$ is terminal, there is only one morphism from $1$ to $1$; thus, the composite morphism $g\circ f$ must be that morphism and because each object of $\mathbf{C}$ must have an identity morphism, $g\circ f = 1_1$. Similarly, $f\circ g = 1_{1'}$. Hence, $1$ and $1'$ are isomorphic. This means that if a category has multiple terminal objects, then these terminal objects must have all properties expressible by morphisms in their category in common.} An element of a $\mathbf{C}$-object $A$ is then defined as the $\mathbf{C}$-morphism $1 \rightarrow A$. Within $\mathbf{Set}$, the terminal objects are the singletons and an element $a$ of a set $A$ can be identified with the morphism $\{a\} \rightarrow A$. Now, from this categorial point of view, if $A$ is the domain of a structure, the ROSRer insists that reference to the elements of this structure does not mean reference to the `internal constitutents' or relata of the structure anymore, but only to terminal objects and corresponding morphisms, which are `external' to the structure itself \cite[\S 2.1]{Bain2011}.

Second, in the categories `sufficiently dissimilar' to $\mathbf{Set}$, where the categorial notion of element (of an object) does not apply (in that the category contains no terminal object for instance), the related worries for ROSR just do not seem to gain any grip. There simply are no elements which could play the role of relata.

According to Bain, both cases then suggest that ``the definition of a structure as an object in a category does not make ineliminable reference to \emph{relata} in the set-theoretic sense.'' (\citeyear[4]{Bain2011}) Indeed, to the extent that everything ``that can be said about sets can be expressed in terms of maps and their composition'', including ``everything about `elements' of sets'' \cite[230]{LawvereandSchanuel1998}, it seems that category theory allows for a notion of structure which, understood as an object in a category, does not explicitly refer to the notion of (internal, primitive) constituents or relata of the structure. So, the first requirement (i) at the end of last section seems to be satisfied.

\section{Throwing out the relations with the relata}
\label{sec:throwing}

However, there are at least two problems with this strategy. First, as Bain himself recognizes, the first case of objects `sufficiently similar' to $\mathbf{Set}$-objects may be subject to the `elimination in name only' charge, so that it may fail to satisfy the second requirement (ii) of \S\ref{sec:peril}. But he argues that in many physically relevant cases the set-theoretic notion of element is not ``essential'' (or just does not apply as in the cases `sufficiently dissimilar' to $\mathbf{Set}$), so that such objections can be avoided (and so requirement (ii) may be satisfied).

This argument can be criticized on various grounds. The precise way in which the set-theoretic notion of element is not ``essential'' in the cases under consideration has to be clarified; Bain's examples from GR and quantum theory are discussed below (this question concerns the requirements (iii) and (iv) of \S\ref{sec:peril}). Even in the cases `sufficiently dissimilar' to $\mathbf{Set}$, it can be argued that a generalized notion of an element of an object $A$ of any category $\mathbf{C}$ can be defined as a $\mathbf{C}$-morphism $X \rightarrow A$, where $X$ is any $\mathbf{C}$-object (not necessarily terminal), so that the notion of elements or constituents of a structure are not truly eliminated, even within the category-theoretic framework and even for categories quite unlike $\mathbf{Set}$.

Secondly, there is a sense in which the categorial ROSRer has thrown out the baby with the bath water: not only the relata of a structure which we take to be exemplified physically have vanished, but also the relations. Let us explain, somewhat technically, how this comes about. Set-theoretically, relations are defined as subsets of Cartesian products, as we have seen above. Now, $n$-ary Cartesian products can be defined in category theory, but they don't exist in all categories.\footnote{A (binary) {\em Cartesian product} of two objects $A$ and $B$ in a category $\mathbf{C}$ is defined to be the ordered triple $\langle A\times B, \pi_1\, \pi_2\rangle$ consisting of an object $A\times B$ of $\mathbf{C}$ and two morphisms
\begin{eqnarray*}
\pi_1: A\times B \rightarrow A,\\
\pi_2: A\times B \rightarrow B,
\end{eqnarray*}
such that for any object $C$ in $\mathbf{C}$ with morphisms $f: C\rightarrow A$ and $g: C\rightarrow B$, there exists a unique morphism $\langle f, g\rangle: C\rightarrow A\times B$ such that
\begin{eqnarray*}
\pi_1\circ \langle f, g\rangle &=& f,\\
\pi_2\circ \langle f, g\rangle &=& g.
\end{eqnarray*}
More intuitively, this means that the following diagram commutes:
\begin{center}
$\xymatrix{
& C \ar[dl]_f \ar@{.>}[d]|-{\langle f, g\rangle} \ar[dr]^g\\
A & \ar[l]^-{\pi_1} A\times B \ar[r]_-{\pi_2} &B}$
\end{center}
This definition can be straightforwardly generalized to define the $n$-fold Cartesian product of any $n$ objects \citep[\S2.1]{Borceux94}. As there is not, in general, a unique morphism $\langle f, g\rangle$ with the required properties, the Cartesian product does not exist in all categories. If it exists, however, the Cartesian product is unique (up to isomorphism).} One might then hope to be able to introduce, in a natural fashion, a categorial correlate of the set-theoretical notion of relations as subsets of Cartesian products in those categories in which the Cartesian product exists. The advantage of this way of introducing relations would be that they would be cleanly attributable to particular objects of a category, which, as above, represent models of the theory. The disadvantage is that in general relations would not exist at all, or only those of a certain arity do. Since the existence of Cartesian products and of terminal objects are independent conditions, we would then face the situation that our best theory may deliver us a category in which either elements of objects, or relations attributable to objects, or both, do not exist. Thus, the debate about whether ROSR or BOSR---if either---is correct turns on the contingencies of theory choice. Arguably, that's a desirable outcome; however, if that is so, as we will see below, it is BOSR which has the upper hand, except in the case the relevant category admits Cartesian products, but no terminal objects. Categories with both terminal objects and Cartesian products (as well as `exponential' objects)\footnote{For a precise definition, cf.~\citet[97f]{maclane98}.} are called {\em Cartesian closed} categories. Some physically important categories are {\em not} Cartesian closed, however; for instance, $\mathbf{Hilb}$ and $\mathbf{nCob}$, to be introduced below, do not have products (for reasons having to do with the non-factorizability of quantum entanglement). 

This shows that not only may there not exist relata, but that there may not be relations, at least not relations as captured by this conceptualization. This may indicate that the central claim of ROSR---that fundamentally, there are only relations, but no relata---becomes hollow, if not plainly false. It may also suggest, however, that the proposed conceptualization fails. An alternative path to introducing relations by carving out the internal, implicit structure of objects of a category is by `concretizing' the category. `Concreteness' is not a property of categories; rather `concretizability' concerns the possibility of equipping a category with additional structure in order to throw a category's structure into sharper relief. A {\em concrete category} is an ordered pair $\langle \mathbf{C}, U\rangle$ where $\mathbf{C}$ is a category and $U: \mathbf{C} \rightarrow \mathbf{Set}$ is a faithful functor from $\mathbf{C}$ to $\mathbf{Set}$. A functor is a morphism of categories, and since $U$ is faithful, a $\mathbf{C}$-morphism $f$ can be identified with its image under $U$, the function $U(f)$.\footnote{More precisely, A {\em functor} $F$ from category $\mathbf{C}$ to category $\mathbf{D}$ is a function that assigns to each $\mathbf{C}$-object $A$ a $\mathbf{D}$-object $F(A)$ and to each $\mathbf{C}$-morphism $f: A \rightarrow B$ a $\mathbf{D}$-morphism $F(f): F(A) \rightarrow F(B)$ such that $F(1_A) = 1_{F(A)}$ and if $g \circ f$ then $F(g \circ f) = F(g) \circ F(f)$. A functor $F: \mathbf{C} \rightarrow \mathbf{D}$ is {\em faithful} just in case for every pair $A, B$ of $\mathbf{C}$-objects and for every pair $f_1, f_2: A\rightarrow B$ of parallel $\mathbf{C}$-morphisms, if $F(f_1) = F(f_2): F(A)\rightarrow F(B)$, then $f_1 = f_2$. Cf.~\citet[\S I]{maclane98}.} Thus, \citet[26]{maclane98} describes a concrete category  as a category $\mathbf{C}$ in which every $\mathbf{C}$-object $A$ ``comes equipped with its `underlying' set [$U(A)$], each [morphism $F:A\rightarrow B$] is an actual function [$U(A)\rightarrow U(B)$ between sets], and composition of [morphisms] is composition of functions.'' Terminologically, we say that a category $\mathbf{C}$ is {\em concretizable} if there exists a faithful functor $U: \mathbf{C}\rightarrow \mathbf{Set}$. Given a concrete category $\langle\mathbf{C}, U\rangle$, the implicit structure of the category can then be explicated by introducing functors $U^n:\mathbf{C}\rightarrow \mathbf{Set}$ for $n\in \mathbb{N}$ such that $U^n(A) = (U(A))^n$. An $n$-ary relation can then be defined as a `subfunctor' of $U^n$.\footnote{A functor $V:\mathbf{C}\rightarrow \mathbf{Set}$ is called a {\em subfunctor} of a functor $U:\mathbf{C}\rightarrow\mathbf{Set}$, symbolically $V\subseteq U$, just in case for all $\mathbf{C}$-objects A, $V(A) \subseteq U(A)$, and for all $\mathbf{C}$-morphisms $f: A\rightarrow B$, $V(f)$ is the restriction of $U(f)$ to $V(A)$.} 

Concretizing a category and studying its implicit structure via its underlying set-theoretical structures and, specifically, the structure of its objects via the relevant subfunctors may be an extremely powerful tool. It does, however, depend on the `exponentiability' of $U(A)$, as is evident in the definition of $n$-ary relations. Since $\mathbf{Set}$ is Cartesian closed, $U(A)^n$ exists. The Cartesian closedness of $\mathbf{Set}$, however, entails that it is also possible to introduce elements of objects in a similar fashion. Thus, if it is legitimate to introduce relations in this way, then it is hard to see why it would be illegitimate to similarly introduce elements of objects. Again, BOSR comes out on top, except perhaps in those cases when the relevant category contains Cartesian products, but no terminal objects.

\section{Categorial and set-theoretical structures}
\label{sec:structures}

The deeper lesson here is that relations, not just relata, are conceptually intimately tied to a set-theoretical understanding of `structure'. In fact, category theory offers a rather different approach to `structure' from the usual set-theoretical one. Given a category $\mathbf{C}$, the structure of a $\mathbf{C}$-object is given by the $\mathbf{C}$-morphisms to and from this object. In category theory, it is by conditions on the web of morphisms of a category that internal structure is imposed on the objects of the category. Whatever these internal structures, they are determined by these conditions only up to isomorphism, as is clear from the definition of isomorphism in Footnote \ref{fn:isom}. Thus, the morphisms only distinguish isomorphism classes of structures. 

An advocate of ROSR might be tempted to feel vindicated by the fact that the web of morphisms thus captures the internal structure of the objects of a category, combined with the assumptions that morphisms can be equated to relations, and that the objects---the relata---are in some sense, precisified above, eliminable. But such a feeling would rest on a confusion: if relational at all, the morphisms are relations between objects, not elements of objects. As characterized above, and as usually in the literature, ROSR is a thesis emphasizing the fundamentality of relations among what is termed `elements' in this context---and not the objects---, at the ontological expense of these elements. Thus, this move cannot save the radical. All this goes to show that the notion of `structure' is generally quite different in category theory from what it is in set theory.\footnote{In fact, a (categorially understood) structure needs not be structured (i.e., contain identifiable `elements')!} Trying to force a categorial understanding of structures into the Procrustean bed of a set-theoretic conception of structure is to miss this deep difference. In this sense, the very debate between radical and balanced OSR is at odds with the way structures get characterized in category theory. 

More fundamentally, the issue here touches the foundational question of the relationship between category theory and set theory. If one can in principle be shown to be reducible to the other, in the sense that it can be formulated purely in terms of the other (while the converse does not obtain) and thus is a special case of it, then that would constitute a sound reason for considering the former less fundamental than the latter. If, for instance, concepts in category theory could be reformulated in entirely set-theoretic terms (but not vice versa), then this would be a strong indication that category theory is less fundamental. This issue arises both concerning the self-sustaining formulation of the basic axioms of category theory, as well as in attempts to articulate the structure of a {\em particular} category.\footnote{Concerning the latter, consider, e.g., the category $\mathbf{Symp}$. Its objects are symplectic manifolds (which encode phase spaces in classical physics) and its morphisms are symplectic transformations. It is not trivial to formulate, in a completely set-theory-free way, all the conditions on the morphisms of the category such that the category's objects are all and only symplectic manifolds and the morphisms exactly turn out to be the symplectic transformations between them.} Arguably, this would also mean that a categorial conception of structure would be reducible to a set-theoretic one. However, if that were the case, the general thrust of Bain's argument would seem to be misdirected, as the argument's whole point is to establish that categorially understood structures are more general and do thus not always afford a translation into set-theoretic terms. To be sure, Bain's argument could be saved even if set theory were more encompassing, but only if supplemented by an argument showing that in the cases of physical interest, category theory grants all the necessary expressive powers {\em and} that the categorial formulations are primary in those cases. 

If, on the other hand, category theory turns out to be more fundamental than set theory, then, as argued just above, there are good reasons to believe that the entire way in which the debate between ROSR and BOSR is cast fails to fully appreciate the fundamental notion of `structure'. To show that there are no relata, i.e.\ no elements of objects, in a relevant category would then suffice to invalidate BOSR; but it wouldn't be enough to vindicate ROSR, since no argument that free-standing relations are fundamental would thereby have been produced.

Even though the fundamentality of category theory may (or may not) be necessary for the argument for free-standing relations to succeed, it is certainly not sufficient. While also connected to the outstanding tasks (iii) and (iv) outlined above (\S\ref{sec:peril}) and discussed below, we would like to add one more point pertaining to foundations here. Suppose we have managed to articulate category theory in terms free of any set-theoretic baggage. As \citet[Appendix]{maclane98} shows, since category theory offers a more general, and hence more permissive, framework than set theory, additional axioms need to be added to those of category theory in order to regain sets as they figure in Zermelo-Fraenkel set theory. First, one adds axioms demanding, e.g., that the categories have terminal objects and Cartesian products. A category satisfying these additional axioms is called {\em topos} (plural {\em topoi}). The category $\mathbf{Set}$ can then be described as a `well-pointed' topos with an axiom of choice and a natural-numbers object, which can be shown to be `weakly consistent' with a version of \textsf{ZF} set theory \citep[291]{maclane98}. If it now turns out that the categories relevant to fundamental physics are all topoi, then the ROSRer must retreat to non-topos categories, and thus to a level of generality, which is simply not called for by fundamental physics, at least as it is known. In this case, however, a moderate is free to refuse imbuing this more general notion of structure with any physical relevance. Alternatively, the ROSRer can downplay the `essentiality' of the elements of the objects of the category, a strategy pursued by Bain, as discussed below.

Be this as it may, Bain recognizes that his argument depends on the relationship between category theory and set theory and acknowledges that showing that category theory is ``more fundamental than set theory'' is the first point on his `must-do' list for ROSR (\citeyear[\S 5]{Bain2011}). However, this is not the place to deal with this controversial question regarding the foundations of mathematics,\footnote{For a recent discussion, see \cite{Pedroso2009} and references therein.} so let us press on.

\section{Radical suggestions from topological quantum field theory?}
\label{sec:TQFT}

Since ROSR is not mathematical structuralism, it is crucial to show that categories and especially those `sufficiently dissimilar' to $\mathbf{Set}$ (\S\ref{sec:bain}) have some physical significance, as demanded by requirements (iii) and (iv) of \S\ref{sec:peril}. More precisely, it has to be shown that the use of categories that do not resemble $\mathbf{Set}$ (in the precise sense that they lack some of the characteristic features of $\mathbf{Set}$) physically amounts of an elimination of the (set-theoretical) notions of objects and relata.

As an example of using category theory to profitable effect in fundamental physics, Bain invokes the precedent of John \cite{Baez2006} and his defence of the physical significance of category theory within the context of developing a quantum theory of gravity. Indeed, \cite{Baez2006} discusses aspects of topological quantum field theory (TQFT), which highlight categorial analogies between GR and quantum theory. The TQFT Baez considers can be understood as a `structure-preserving' map (i.e., a functor) from the category $\mathbf{nCob}$, whose objects are compact oriented $(n-1)$-manifolds and whose morphisms are compact oriented $n$-manifolds with a boundary that is composed by the corresponding dom and cod objects, to a category $\mathbf{Hilb}$, whose objects are Hilbert spaces and whose morphisms are bounded linear operators. The categories $\mathbf{nCob}$ and $\mathbf{Hilb}$ have obvious physical significance in GR and quantum theory, respectively.\footnote{Even though a number of technical and conceptual details remain to be clarified, such as whether in the category $\mathbf{Hilb}$, for instance, any Hilbert space should be considered, or only separable ones, and whether any bounded linear operator is a morphism, or only essentially self-adjoint ones, etc.} Leaving aside the details, Baez's main point is to emphasize the fact that $\mathbf{Hilb}$ and $\mathbf{nCob}$ share category-theoretic properties that $\mathbf{Set}$ does not possess; and since these properties precisely encode fundamental quantum features (such as non-separability and no-cloning) as well as define the spacetime counterparts to these quantum features (through the TQFT), he argues, speculatively, that the categoric point of view `outside' of $\mathbf{Set}$ might be fruitful for the development of a theory of quantum gravity.\footnote{\label{fn:ncobhilb}Specifically, the differences between $\mathbf{nCob}$ and $\mathbf{Hilb}$, on the one hand, and $\mathbf{Set}$, on the other, are threefold. First, the former, but not the latter, are so-called `monoidal' categories, which admit a tensor product instead of Cartesian products. This means that the structure of `limits' differs in the two cases. Second, the former have additional structure that the latter lacks: each of their morphisms has an `adjoint' morphism. Third, in Bain's own words, ``the objects of $\mathbf{nCob}$ and $\mathbf{Hilb}$ cannot be structured sets, insofar as their morphisms are not simply functions that preserve the relevant set-theoretic notion of structure associated with them... Set-theoretically, the functions that preserve the structure of a Hilbert space are unitary operators that preserve the inner-product. The morphisms of $\mathbf{Hilb}$ in contrast are general bounded linear operators that do not necessarily have to be unitary.'' (10) (Instead, the inner product is defined via an adjoint operation.) It should be noted that because these operators do not have to be unitary the category $\mathbf{Hilb}$ offers a generalization of our usually unitary quantum theories. This generalization can ultimately only possibly underwrite ROSR if it is physically called for and hence justified. The suggestive analogy to $\mathbf{nCob}$ seem to arise precisely because unitarity is dropped.}

However, despite their dissimilarities with $\mathbf{Set}$, $\mathbf{nCob}$ and $\mathbf{Hilb}$ do allow for a notion of element of an object (so they belong to the first class described in \S\ref{sec:bain}). But Bain argues that the elements of the objects in $\mathbf{nCob}$ (manifold points) and in $\mathbf{Hilb}$ (Hilbert space vectors) are not ``essential'' since the objects in $\mathbf{nCob}$ and $\mathbf{Hilb}$ cannot be considered as structured sets (i.e.\ sets with a structure, such as the ordered pair $\langle A, R\rangle$ in \S\ref{sec:peril}) in the categorial context, since the morphisms in $\mathbf{nCob}$ and $\mathbf{Hilb}$ are not `structure-preserving' enough. From the categorial point of view, only what is preserved by the morphisms is part of the relevant structure. Thus, it is not possible to obtain the categories $\mathbf{nCob}$ and $\mathbf{Hilb}$ from the category of structured sets by imposing further conditions on the morphisms. Furthermore, the states of a quantum system---the physically relevant entities in $\mathbf{Hilb}$---do not correspond to the elements of the relevant $\mathbf{Hilb}$-object (different vectors can correspond to the same state); in this sense and in the context of the considered TQFT, the categorial elements of the $\mathbf{Hilb}$-objects are therefore not ``essential'' entities.

As suggestive as they may be, these categoric considerations should however not make us lose sight of the following facts. The categoric dissimilarities with $\mathbf{Set}$ might well be fruitful to develop a theory of quantum gravity, in particular in using TQFT as a tool or a toy model.  However, these dissimilarities do not entail that there are no physical objects and relata at all. Although we will not argue the point here---the reader is invited to consult \cite{EsfeldandLam2011} and \cite{LamandEsfeld2012} in the structuralist context, quantum mechanics and quantum field theory are about quantum (field) systems whose nature is clearly radically different from everyday classical objects but which should nevertheless be considered as genuine physical objects on their own rights, at least in the sense in which they provide an account for the measurements we make. Bain acknowledges this point and even considers it a deep concern (\citeyear[\S 5]{Bain2011}). The category-theoretic framework does not eliminate references to them and does not provide, to date, any physical ground for such elimination, but it does provide useful mathematical tools, allowing in particular to build mathematical `bridges'---the functors---between different spacetimes and quantum (field) theory, thereby encoding (to some extent) the dynamical nature of spacetime demanded by GR. This is precisely the aim of TQFT.\footnote{More recently, but in the same spirit, algebraic quantum field theory has been defined as a functor between the relevant categories, see \cite{BrunettiFredenhagenandVerch2003}.}

\section{Sheaves of Einstein algebras as radical structures?}
\label{sec:einalg}

Bain's second, and main, example of categorial elimination of the notion of physical object concerns spacetime points within some categoric and sheaf-theoretic extension of GR. The general idea is to generalize the algebraic formulation of GR in terms of Einstein algebras using sheaves of Einstein algebras (and working within the corresponding category of sheaves).\footnote{Very roughly, a sheaf can be understood as a generalization of a fibre bundle, where the `fibres'---here Einstein algebras---are assigned to open sets of the `base space'. See below for a more detailed treatment.} The algebraic formulation of GR relies on the full equivalence of, on the one hand, the usual (geometric) definition of a smooth differentiable manifold $M$ in terms of a set of points with a topology and a smooth differential structure with, on the other hand, the alternative (algebraic) definition of $M$ using only the algebraic structure of the (commutative) ring $C^{\infty}(M)$ formed by the set of the smooth real functions on $M$ (under pointwise addition and multiplication; indeed $C^{\infty}(M)$ is an algebra). Tensorial calculus and equations on $M$---such as the Einstein field equations---can be defined and carried out in purely algebraic terms; an {\em Einstein algebra}, then, is defined as a linear algebra (without explicit reference to the smooth differentiable manifold $M$) that allows the relevant algebraic constructions (corresponding to a Lorentz metric, a covariant derivative, etc.) such that the algebraic Einstein field equations are satisfied. The original formulation of GR in such algebraic terms, which is completely equivalent to the standard geometric one, is due to \cite{Geroch1972}. 

The relata of the spacetime models in the standard tensorial formulation of GR are the points of the manifold $M$. We concur with Bain that the explicit reference to manifold points is eliminated; it is, however, replaced with a reference to `maximal ideals', which become central in the recovery of the differentiable manifold from the algebraic structure. A {\em maximal ideal} of a commutative algebra $\mathcal{A}$ is the largest proper subset---indeed a subgroup of the additive group---of $\mathcal{A}$ closed under multiplication by any element of $\mathcal{A}$. The maximal ideal of $C^{\infty}(M)$ that corresponds to a point $p \in M$ is the set of all functions vanishing at $p$. This correspondence, it should be noted, is 1-1, i.e., there is a bijective map between the points of a differentiable manifold and the maximal ideals of $C^{\infty}(M)$.\footnote{As Elias Zafiris has pointed out to us, this is a special case of the Gelfand representation theorem. As \citet[\S3]{Bain2011} notes, this can be regarded as a ``demonstration that tensor models and [Einstein algebra] models belong to the same isomorphism class of structured sets, and hence encode the same structure.'' As mentioned above, some structural realists have taken this as an indication that it is this structure, rather than the differentiable manifold-cum-metric, that we ought to be realists about. However, if this is an argument in favour of OSR, then it seems to be as much an argument for BOSR as for ROSR.} This isomorphism immediately raises the worry that the elimination of manifold points was ``in name only''. In order to appease this concern, \citet[6]{Bain2011} argues that ``at least for some solutions to the Einstein equations, the move to the Einstein algebra formalism is in fact a move that non-trivially eliminates manifold points (and their [Einstein algebra] correlates), but retains differentiable structure.'' Before we consider this argument, let us emphasize that the mere non-trivial elimination of manifold points (or their correlates) does not suffice to give ROSR the upper hand against BOSR. What must be shown in order to establish {\em that}, instead, is that {\em any} relata whatsoever are non-trivially eliminated from the formulation of the theory; BOSR does not include or entail a commitment to any specific set of relata such as manifold points. We shall return to this point below. 

Thus, the models of the algebraic formulation of GR are structured and thus don't eliminate reference to relata; they don't even truly eliminate reference to manifold points, which are merely re-labelled `maximal ideals' in the algebraic context. This is where the generalization of algebraic GR based on {\em sheaves} of Einstein algebras enters the stage. Algebraic GR (but not standard GR in its tensorial garb) can be extended to include `completed' spacetimes with base manifolds $M' = M \cup \partial M$, where $\partial M$ is a boundary space attached to $M$. This extension can be achieved, in a fairly unified way to be precisified, by using sheaf-theoretic methods. The physical motivation for such a generalization of GR is that it provides a formalism that may be useful for investigating the nature of spacetime singularities.\footnote{Bain refers to the work of Michael Heller and co-workers, see for instance \cite{HellerandSasin1995}; they have recently further generalized this algebraic approach to GR using non-commutative methods, the aim being to provide a framework for a theory of quantum gravity, see for instance \cite{HellerandPysiakandSasin2005}.} Now the claim that spacetime points are eliminated or play ``no essential role'' relies on the fact that the algebraic counterparts of manifold points (maximal ideals of the Einstein algebras) may not be defined within the sheaf-theoretic and categoric framework: indeed, the elements of the objects of the category of `sheaves of Einstein algebras' may not be defined at all.

Let us look at all this in some detail. In a first approach, sheaves can be defined as functors from the category of open sets of a topological space to the category of sets satisfying certain conditions (see below for precise definitions). In the case at hand, the Einstein algebra $\langle C, g\rangle$ defined on $M$ is replaced with a `sheaf' of Einstein algebras $\langle \mathcal{E}, g\rangle$ defined on $M'$ where the ring $C \cong C^{\infty} (M)$ is replaced with the `sheaf' $\mathcal{E} \cong C^{\infty} (M \cup \partial M)$. Intuitively, a sheaf of Einstein algebras is a ``collection of Einstein algebras indexed by the open regions on $M$ induced by the topology on $M'$.'' \citet[7]{Bain2011} The claimed advantage of bringing to bear this entire abstract machinery (cf.\ below) is the physically salient unification it allegedly brings, e.g.\ in investigations of singularities. This idea of a unified treatment of spacetimes with and without boundaries is supported by the recognition that $\langle C, g\rangle$ and $\langle \mathcal{E}, g\rangle$ can be subsumed into the same category---the category of sheaves of Einstein algebras on $M'$, or, more precisely, the category of `Einstein structured spaces'.\footnote{An {\em Einstein structured space} is a pair $(M, \mathcal{E})$, where $M$ is a topological space, e.g. the differentiable manifold in the standard geometric formulation of GR, and $\mathcal{E}$ is a sheaf of Einstein algebras over $M$ satisfying certain conditions with respect to the composition of smooth functions (see \citealp{HellerandSasin1995}).} Such a unified categorial treatment is not available in the standard tensorial formalism.

Before we proceed with Bain's argument, let us note a grave concern. We agree with \citet[3657]{HellerandSasin1995} that Einstein structured spaces are ``more general than the usual smooth manifold since [they contain] as [their] intrinsic elements all space-time singularities which do not violate the property of $M$ being of constant differential dimension.'' However, such a generalization is only relevant if the spacetimes with boundaries it incorporates have physical significance. The justification for this significance, unfortunately, is rather mute.\footnote{It can be argued that the sheaf-theoretic (and the non-commutative) approach to space-time singularities highlights their fundamental global features, see \cite{HellerandSasin1994} and the discussion in \cite{Lam2007}; in this sense, the singular boundary constructions within this sheaf-theoretic context may constitute useful conceptual tools, whose possible physical significance depends on the success of the future developments in this approach. In any case, the global aspects of space-time singularities do not entail `free-standing' structures (see the discussion below).} With the exception of the idealized case of glueing a conformal boundary to an asymptotically flat spacetime via a Penrose construction, boundary construction seem to lack physical significance, at least in the context of GR.\footnote{We wish to thank Erik Curiel for drawing our attention to this point and refer the reader to his \citeyear{Curiel1999} for detailed arguments in support of this claim, at least as it pertains to the case of singular boundaries.} In this sense, the category of completed spacetimes simply seems to be the wrong category as far as GR is concerned. The claimed advantage of the more general framework of Einstein structured spacetimes is that it classifies the general-relativistic spacetimes and the additional ones with boundaries as belonging to the same category \cite[\S 3.3]{Bain2011}. In the light of the questionable physical relevance of these boundary constructions, we fail to see why the lonely fact that these rather different cases are subsumed under a single category in itself constitutes a decisive advantage. Thus, our requirement (iv) of Section \ref{sec:peril} is certainly not satisfied. In the absence of such an advantage, the best the ROSRer can hope for is to show that in some physical contexts involving boundary constructions, e.g.\ in the study of certain types of singularities, we may beneficially apply theories which do not ultimately rely on the existence of relata. Provided the argument gets traction at all (there are good reasons to think that it does not, see below), this would be an interesting result, no doubt; but it is a far cry from offering a sound argument to the effect that our most encompassing fundamental theories ought to be interpreted as making an ontological commitment to `free-standing' structures sans relata.

Of course, an advocate of ROSR could accept that recasting GR in terms of the category of Einstein structured spaces would yield no relevant new physics, yet insist that the reformulated theory was the metaphysically more perspicuous account of the empirical phenomena identically saved by the two formulations.\footnote{We thank Oliver Pooley for pressing us on that point.} We certainly accept the possibility of this strategy. In order to amount to more than mere assertion, the insistence would have to be accompanied by an argument for the claimed perspicuity. Such an argument would have to make the contrastive case for ROSR {\em vis-\`a-vis BOSR}, in particular, to be effective in the present context. 

To return to Bain's argument, more generally---and readers uninterested in the technical details are invited to skip this paragraph---, a sheaf is a `pre-sheaf' which satisfies certain additional conditions. Following a standard textbook \citep[\S1.1]{Tennison1975} on sheaf theory quite closely, a {\em pre-sheaf} $F$ of sets on a topological space $X$ is defined by two conditions: (i) for every open set $U$ of $X$, a set $F(U)$ (called the set of `sections' of $F$ over $U$) is given, and (ii) for every pair of open sets $V\subseteq U$ of $X$, a {\em restriction map} $\rho_{UV}: F(U) \rightarrow F(V)$ such that (a) for all $U$, $\rho_{UU}$ is the identity on $U$, and (b) whenever $W\subseteq V\subseteq U$ for open sets $U, V$, and $W$, $\rho_{UW} = \rho_{VW} \circ \rho_{UV}$, meaning that 
\[
\xymatrix{
F(U) \ar[dr] \ar[rr] & & F(W) \\
& F(V) \ar[ur]}
\] 
commutes. A pre-sheaf $F$ of sets over $X$ satisfying the two conditions (M) and (G) is called a {\em sheaf} of sets. The conditions are as follows \citep[\S2.1]{Tennison1975} :
\begin{enumerate}
\item[(M)] If $U$ is an open set of $X$ and $U = \cup_{\lambda \in \Lambda} U_\lambda$ an open covering of $U$, i.e.\ each $U_\lambda$ is open in $X$, and $s, s' \in F(U)$ are two sections of $F$ such that 
\[
\forall \lambda \in \Lambda, \rho_{UU_\lambda} (s) = \rho_{UU_\lambda} (s'),
\]
then $s = s'$. (The satisfaction of this local identity condition defines `monopresheaves'.)
\item[(G)] If $U$ is an open set of $X$ and $U = \cup_{\lambda \in \Lambda} U_\lambda$ an open covering of $U$, and if we are given a family $(s_\lambda)_{\lambda\in \Lambda}$ of sections of $F$ with $\forall \lambda\in\Lambda, s_\lambda \in F(U_\lambda)$ such that
\[
\forall\lambda \in \Lambda, \rho_{U_\lambda U_{\lambda\mu}} (s_\lambda) = \rho_{U_\mu U_{\lambda\mu}} (s_\lambda)
\]
where $U_{\lambda\mu} \doteq U_\lambda \cap U_\mu$, then there exists an $s\in F(U)$ such that
\[
\forall \lambda \in \Lambda, \rho_{UU_\lambda} (s) = s_\lambda.
\]
\end{enumerate}
\citet[15]{Tennison1975} paraphrases the `glueing' condition (G) as demanding that ``if a system $(s_\lambda)$ is given on a covering and is consistent on all overlaps, then it comes from a section over all of $U$.'' The resulting section $s$, whose existence is guaranteed by (G), is `glued together' from the sections $s_\lambda$. Combined (M) and (G) assert that compatible sections can be uniquely patched together. Generally, sheaves, although they can be understood as functors between categories, form their own category---the category of sheaves $\mathbf{Sh(X)}$ on $X$ (for the category of sets). The objects of this category are the sheaves, and the morphisms are morphisms between sheaves defined for each open set of $X$ compatible with the restrictions $\rho$. 

The question relevant to the present essay is whether or not the objects of this category, i.e., the sheaves, have elements.\footnote{We wish to thank Elias Zafiris for corresponding on this point and his many helpful clarifications. On the physical interpretation of sheaf theory, see \citet{Zafiris2009}.} The definitions guarantee that a sheaf has always local sections; however, sections locally defined over open regions of the underlying topological space only are not in general combinable to `global' sections defined over the whole manifold. In general, therefore, a sheaf does not have global sections. Whether or not, and to what extent, local sections can be extended continuously to global ones is the topic of an entire theory called `sheaf cohomology theory'. Specifically to the case at hand, sections of a sheaf of Einstein algebras are, as for other sheaves, defined over open sets constituted by the points of the underlying manifold $M$ (or $M'$, as the case may be) or, equivalently, of the maximal ideals of the relevant Einstein algebra.\footnote{This means that these sections are animals of quite distinct a character from that of the maximal ideals. It is thus not the case, as \citet[9]{Bain2011} suggests, that ``[a] section of a sheaf of Einstein algebras over an open region $U$ of $M$ is an element of the Einstein algebra assigned to $U$; namely, it is a maximal ideal of that algebra.'' Intuitively, sections are a bit like `functions' on $M$, while maximal ideals correspond to the points of $M$, i.e., the former are the sort of things which are predicated of the latter sort of things. We are grateful to Elias Zafiris to drawing our attention to this point.} For the relevant sheaves, cohomology theory tells us that whether or not global sections exist depends on the topology of the underlying manifold. In general, however, they do not. 

The next step in Bain's argument is to equate elements of the sheaves, i.e., the relevant candidate {\em relata}, the morphisms from the terminal object to the objects of the category of structured sets as identified above, with the sheaf's global sections. The absence of global sections, he claims, indicates an instance of the second type of categories---`sufficiently dissimilar' to $\mathbf{Set}$---as described in Section \ref{sec:bain}. If only global sections could possibly play the role of relata, then the fundamental existence of relata would indeed depend on the existence of global sections of the relevant sheaf, and thus, in turn, on the topological structure of the underlying manifold. We agree with Bain that if a sheaf does possess global sections, these can straightforwardly be identified as its elements. However, even if global sections do not exist, there remains an abundance of local sections and we see no reason why it would be illegitimate to interpret these local sections as local elements, except perhaps that the interpretive task of working out the structure of the sheaf becomes richer. Once we have `local' elements, furthermore, these can surely play the role of relata, however local, that stand in (local) relations to one another. It is just that the global structure will then be constituted not by a simple set of relata and the relations they exemplify, but rather by a family of local relational structures. If the world turned out to be like that, there is no reason why the BOSRer could not consider this an interesting discovery about the fundamental constitution of reality perfectly compatible with her stance. 

But even if it is granted that global sections are required for there to be fundamental relata, it eludes us why the detour through sheaf theory was necessary to make what we take to be the intended point. The fact that the notion of spacetime point---albeit not the notion of {\em relata}---disappears from such sheaf-theoretic models is not controversial (although, as we have argued, the physical relevance of this mathematical fact of course depends on the physical significance of the model). Sections, global or local, do not correspond to maximal ideals and hence not to manifold points. Indeed, the notion disappears in the non-commutative spaces associated with non-commutative (Einstein) algebras as well---non-commutative algebras in general have no maximal ideals. We recognize that this requires another generalization of GR, a generalization which would also have to be justified on the basis of physical significance. However, it goes to show that there is absolutely no need of categorial tools in order to reach the conclusion that spacetime points may well not be fundamental. Moreover, the disappearance of the notion of spacetime points at some more fundamental level than classical GR---perhaps at the quantum gravitational level---is suggested by other approaches to quantum gravity, too. Indeed, spacetime as such may not be part of the ontology of such theories.\footnote{For an argument to this effect, see \citet{HuggettWuthrich2012}.}

What is much more controversial, however, is the move from the (possible) elimination of the spacetime points at some fundamental level (described by sheaf-theoretic tools) to the conclusion that spacetime is a `free-standing' structure \`a la ROSR. According to Bain, this conclusion follows since the spacetime points are the ``\emph{relata} on which spatiotemporal properties are predicated'' (\citeyear[6]{Bain2011}). But one can object that while the sheaf-theoretic (or indeed non-commutative) generalization of GR eliminates reference to spacetime points, it may, however, refer to other physical objects and relata, as described within its formalism, to which (possibly spatio-temporal) properties are predicated. This is not the place to address the speculative questions of what these objects as well as their physical and empirical significance could be (the scalar fields constituting the Einstein algebras are an option), but the following point should be clear: in order to deal BOSR a serious, and perhaps fatal, blow, it has to be shown that not only reference to spacetime points can be eliminated but reference to any physical object or relata {\em tout court}. But the sheaf-theoretic extension of GR does not achieve this. 

In rather general terms, \citet[13, and 9]{Bain2011} seems to think that whatever the relata may be, they have to be {\em localized}, or at least {\em localizable}, individuals or entities.\footnote{This may explain his focus on manifold points, or vice versa. It may also be the reason why he ends by asserting that ``the Einstein algebra formulation of general relativity supports an ontology of structure devoid of objects.'' (14) We take ourselves to have shown that this is not so---and indeed he has admitted as much, e.g.\ on page 6.} Apart from notorious difficulties of characterizing `locality' in a sensible way in GR (although the draconic restriction to manifold {\em points} evades these difficulties), a demand that genuine relata be localized is in no way part of BOSR's commitment. And that's a good thing, too: after all, locality, or localizability, is clearly a spatial or spatio-temporal concept, and space or spacetime may not survive, at least not qua fundamental existent, the move from a classical theory of gravity to a more fundamental quantum theory of gravity \citep{HuggettWuthrich2012}. Furthermore, even at the classical level, the emphasis on local relata seems ill justified; after all, we have seen that the maximal ideals of the algebraic formulation of GR correspond to the manifold points of the usual tensorial language. And while the manifold points of standard GR, if anything, are obvious candidates for localized relata, the maximal ideals rely on a more `global' characterization of the structure of a model as is essential in algebraic GR. The equivalence of standard GR with algebraic GR, and the straightforward correspondence between manifold points and maximal ideals, suggests that the `globality' or `locality' of the relata may not be essential to GR and to the description of its models.\footnote{It is perhaps not surprising, then, that global features of spacetimes like singularities and asymptotic properties can be well characterized in algebraic GR (see \citealp{Lam2007, Lam2008}), and that Bain's argument, if it gets traction at all, really only gets some grip in this case. A similar point obtains regarding TQFT, according to which there are no local degrees of freedom.}

\section{Conclusions}
\label{sec:conclusions}

We have argued that category theory does not provide a relevantly more hospitable environment than set theory for ROSR and its controversial ontological thesis of `free-standing' relations altogether devoid of relata, and hence of any objects whatsoever. One of our central points is that the notions of relata and relations, to the extent that they are physically meaningful, are conceptually intimately tied to a set-theoretical understanding of `structure'; indeed, \emph{both} set-theoretic notions of relata and relations may vanish within the framework of the category-theoretic understanding of `structure'. Moreover, the categorial $\mathbf{C}$-morphisms, which constitute the categorial structure, hold among categorial $\mathbf{C}$-objects---not among their elements; however, these latter represent the fundamental physical objects ROSR wants to eliminate in favour of the physical relations in which they stand represented by relations among elements of $\mathbf{C}$-objects, not $\mathbf{C}$-morphisms among $\mathbf{C}$-objects. So, $\mathbf{C}$-morphisms among $\mathbf{C}$-objects are not the right candidates---the right structures---for ROSR to be realist about.\footnote{Or maybe they are; but in that case, the philosophical gloss ought to be rather different from the usual ROSR rhetoric.}

Beyond these rather abstract considerations, we have also critically discussed two `concrete' physical examples of alleged category-theoretic support to ROSR recently put forward by \citet{Bain2011}: we have argued that the categorial tools applied to TQFT and to the algebraic (sheaf-theoretic) generalisation of GR do not warrant the ROSR claim that these physical theories describe `object-free' structures. In particular, the fact that {\em standard} fundamental physical objects or relata (such as spacetime points) might not be part of the ontology of these candidate fundamental physical theories in their categorial formulation does not imply that there are no physical objects or relata at all. 

In drawing radical lessons from contemporary fundamental physics, ROSR aims to contribute to the naturalistic approach to metaphysics; within this framework, one should welcome metaphysical revisions that are physically well-motivated. However, whatever the ontology suggested by fundamental physics, one must be able to formulate it in a coherent and precise way. It is doubtful whether such a formulation of ROSR with its rejection of objects can obtain in the set-theoretic framework; we have shown in this paper that moving to category theory does not help. Furthermore, by rejecting objects and relata, it can be argued that ROSR merely leaves the job of naturalistic structuralism unfinished, as the broad structuralist framework suggested by fundamental physical features (e.g. quantum non-locality) for the `objects' described by contemporary fundamental physics (e.g. quantum fields) remains to be explicitly articulated.

\bibliographystyle{plainnat}
\bibliography{Ref}

\begin{thebibliography}{38}
\providecommand{\natexlab}[1]{#1}
\providecommand{\url}[1]{\texttt{#1}}
\expandafter\ifx\csname urlstyle\endcsname\relax
  \providecommand{\doi}[1]{doi: #1}\else
  \providecommand{\doi}{doi: \begingroup \urlstyle{rm}\Url}\fi

\bibitem[Ainsworth(2010)]{Ainsworth2010}
P.~M. Ainsworth.
\newblock What is ontic structural realism?
\newblock \emph{Studies in History and Philosophy of Modern Physics},
  41:\penalty0 50--57, 2010.

\bibitem[Baez(2006)]{Baez2006}
J.~Baez.
\newblock {Quantum quandaries: A category-theoretic perspective}.
\newblock In D.~Rickles and S.~French, editors, \emph{{The Structural
  Foundations of Quantum Gravity}}, pages 240--265. Oxford University Press,
  Oxford, 2006.

\bibitem[Bain(2013)]{Bain2011}
J.~Bain.
\newblock Category-theoretic structure and radical ontic structural realism.
\newblock \emph{Synthese}, 190:\penalty0 1621--1635, 2013.

\bibitem[Borceux(1994)]{Borceux94}
F.~Borceux.
\newblock \emph{Handbook of Categorical Algebra 1: Basic Category Theory}.
\newblock Cambridge University Press, Cambridge, 1994.

\bibitem[Brunetti et~al.(2003)Brunetti, Fredenhagen, and
  Verch]{BrunettiFredenhagenandVerch2003}
R.~Brunetti, K.~Fredenhagen, and R.~Verch.
\newblock The generally covariant locality principle---a new paradigm for local
  quantum field theory.
\newblock \emph{Communications in Mathematical Physics}, 237:\penalty0 31--68,
  2003.

\bibitem[Chakravartty(2003)]{Chakravartty2003}
A.~Chakravartty.
\newblock {The structuralist conception of objects}.
\newblock \emph{Philosophy of Science}, 70:\penalty0 867--878, 2003.

\bibitem[Curiel(1999)]{Curiel1999}
E.~Curiel.
\newblock The analysis of singular spacetimes.
\newblock \emph{Philosophy of Science}, 66:\penalty0 119--145, 1999.

\bibitem[Esfeld and Lam(2008)]{EsfeldandLam2008}
M.~Esfeld and V.~Lam.
\newblock {Moderate structural realism about space-time}.
\newblock \emph{Synthese}, 160:\penalty0 27--46, 2008.

\bibitem[Esfeld and Lam(2011)]{EsfeldandLam2011}
M.~Esfeld and V.~Lam.
\newblock {Ontic structural realism as a metaphysics of objects}.
\newblock In A.~Bokulich and P.~Bokulich, editors, \emph{Scientific
  structuralism}, pages 143--159. Springer, Dordrecht, 2011.

\bibitem[French(1998)]{French1998}
S.~French.
\newblock On the withering away of physical objects.
\newblock In E.~Castellani, editor, \emph{Interpreting Bodies: Classical and
  Quantum Objects in Modern Physics}, pages 93--113. Princeton University
  Press, Princeton, 1998.

\bibitem[French(2006)]{French2006}
S.~French.
\newblock {Structure as a weapon of the realist}.
\newblock \emph{Proceedings of the Aristotelian Society}, 106:\penalty0
  169--187, 2006.

\bibitem[French(2010)]{French2010}
S.~French.
\newblock {The interdependence of structure, objects and dependence}.
\newblock \emph{Synthese}, 175:\penalty0 89--109, 2010.

\bibitem[French(2012)]{french2012}
S.~French.
\newblock The presentation of objects and the representation of structure.
\newblock In Elaine Landry and Dean Rickles, editors, \emph{Structure, Object,
  and Causality: Proceedings of the Banff Workshop on Structural Realism},
  University of Western Ontario Series in Philosophy of Science, pages 3--28.
  Springer, Dordrecht, 2012.

\bibitem[French and Ladyman(2003)]{FrenchandLadyman2003}
S.~French and J.~Ladyman.
\newblock Remodelling structural realism: Quantum physics and the metaphysics
  of structure.
\newblock \emph{Synthese}, 136\penalty0 (1):\penalty0 31--56, 2003.

\bibitem[Geroch(1972)]{Geroch1972}
R.~Geroch.
\newblock {Einstein Algebras}.
\newblock \emph{Communications of Mathematical Physics}, 26:\penalty0 271--275,
  1972.

\bibitem[Goldblatt(1984)]{Goldblatt1984}
R.~Goldblatt.
\newblock \emph{{Topoi}}.
\newblock Elsevier, Amsterdam, 1984.

\bibitem[Heller and Sasin(1994)]{HellerandSasin1994}
M.~Heller and W.~Sasin.
\newblock The structure of the $b$-completion of space-time.
\newblock \emph{General Relativity and Gravitation}, 26:\penalty0 797--811,
  1994.

\bibitem[Heller and Sasin(1995)]{HellerandSasin1995}
M.~Heller and W.~Sasin.
\newblock Structured spaces and their application to relativistic physics.
\newblock \emph{Journal of Mathematical Physics}, 36:\penalty0 3644--3662,
  1995.

\bibitem[Heller et~al.(2005)Heller, Pysiak, and
  Sasin]{HellerandPysiakandSasin2005}
M.~Heller, L.~Pysiak, and W.~Sasin.
\newblock {Noncommutative unification of general relativity and quantum
  mechanics}.
\newblock \emph{Journal of Mathematical Physics}, 46:\penalty0 122501, 2005.

\bibitem[Huggett and W\"uthrich(forthcoming)]{HuggettWuthrich2012}
N.~Huggett and C.~W\"uthrich.
\newblock Emergent spacetime and empirical (in)coherence.
\newblock \emph{Studies in History and Philosophy of Modern Physics},
  forthcoming.

\bibitem[Ladyman(2007)]{Ladyman2007}
J.~Ladyman.
\newblock On the identity and diversity of objects in a structure.
\newblock \emph{Proceedings of the Aristotelian Society Supplementary Volume},
  81\penalty0 (1):\penalty0 45--61, 2007.

\bibitem[Ladyman et~al.(2007)Ladyman, Ross, Spurett, and
  Collier]{LadymanRossSpurrettandCollier2007}
J.~Ladyman, D.~Ross, D.~Spurett, and J.~Collier.
\newblock \emph{{Every Thing Must Go: Metaphysics Naturalized}}.
\newblock Oxford University Press, Oxford, 2007.

\bibitem[Lam(2007)]{Lam2007}
V.~Lam.
\newblock {The singular nature of spacetime}.
\newblock \emph{Philosophy of Science}, 74:\penalty0 712--723, 2007.

\bibitem[Lam(2008)]{Lam2008}
V.~Lam.
\newblock {Structural aspects of the singular feature of space-time}.
\newblock In D.~Dieks, editor, \emph{Ontology of Spacetime. Philosophy and
  Foundations of Physics Series. Vol.2}, pages 111--131. Elsevier, Amsterdam,
  2008.

\bibitem[Lam and Esfeld(2012)]{LamandEsfeld2012}
V.~Lam and M.~Esfeld.
\newblock The structural metaphysics of quantum theory and general relativity.
\newblock \emph{Journal for General Philosophy of Science}, 43:\penalty0
  243--258, 2012.

\bibitem[Landry(2007)]{Landry2007}
E.~Landry.
\newblock {Shared structure need not be shared set-structure}.
\newblock \emph{Synthese}, 158:\penalty0 1--17, 2007.

\bibitem[Lawvere and Schanuel(1998)]{LawvereandSchanuel1998}
F.~W. Lawvere and S.~H. Schanuel.
\newblock \emph{{Conceptual Mathematics}}.
\newblock Cambridge University Press, Cambridge, 1998.

\bibitem[Mac~Lane(1998)]{maclane98}
S.~Mac~Lane.
\newblock \emph{Categories for the Working Mathematician}.
\newblock Springer, New York, 2nd edition, 1998.

\bibitem[Muller(2010)]{Muller2010}
F.~A. Muller.
\newblock The characterisation of structure: Definition versus axiomatisation.
\newblock In F.~Stadler, editor, \emph{The Present Situation in the Philosophy
  of Science}, pages 399--416. Springer, 2010.

\bibitem[Muller(2011)]{Muller2011}
F.~A. Muller.
\newblock How to defeat {W\"uthrich's} abysmal embarrassment argument against
  space-time structuralism.
\newblock \emph{Philosophy of Science}, 78:\penalty0 1046--1057, 2011.

\bibitem[Muller and Saunders(2008)]{MullerandSaunders2008}
F.~A. Muller and S.~Saunders.
\newblock Discerning fermions.
\newblock \emph{British Journal for the Philosophy of Science}, 59:\penalty0
  499--548, 2008.

\bibitem[Muller and Seevinck(2009)]{MullerandSeevinck2009}
F.~A. Muller and M.~Seevinck.
\newblock Discerning elementary particles.
\newblock \emph{Philosophy of Science}, 76:\penalty0 179--200, 2009.

\bibitem[Pedroso(2009)]{Pedroso2009}
M.~Pedroso.
\newblock {On three arguments against categorical structuralism}.
\newblock \emph{Synthese}, 170:\penalty0 21--31, 2009.

\bibitem[Saunders(2003)]{Saunders2003b}
S.~Saunders.
\newblock {Structural realism, again}.
\newblock \emph{Synthese}, 136:\penalty0 127--133, 2003.

\bibitem[Tennison(1975)]{Tennison1975}
B.~R. Tennison.
\newblock \emph{Sheaf Theory}, volume~20 of \emph{London Mathematical Society
  Lecture Note Series}.
\newblock Cambridge University Press, Cambridge, 1975.

\bibitem[W\"uthrich(2009)]{Wuthrich2009}
C.~W\"uthrich.
\newblock Challenging the spacetime structuralist.
\newblock \emph{Philosophy of Science}, 76:\penalty0 1039--1051, 2009.

\bibitem[W\"uthrich(2012)]{Wuthrich2011}
C.~W\"uthrich.
\newblock {The structure of causal sets}.
\newblock \emph{Journal for General Philosophy of Science}, 43:\penalty0
  223--241, 2012.

\bibitem[Zafiris(2009)]{Zafiris2009}
E.~Zafiris.
\newblock A sheaf-theoretic topos model of the physical continuum and its
  cohomological observable dynamics.
\newblock \emph{International Journal of General Systems}, 38:\penalty0 1--27,
  2009.

\end{thebibliography}

\end{document}